\documentclass[aps,prl,twocolumn]{revtex4-1}
\usepackage{graphicx}% Include figure files
\usepackage{dcolumn}% Align table columns on decimal point
\usepackage{bm}% bold math
\usepackage{hyperref}
\usepackage{subfigure}

\begin{document}

\title{Reply to Flambaum and Porsev comment on ``21 cm radiation  - a new probe of variation in the fine structure constant''}
\author{Rishi Khatri}
\email{rkhatri2@illinois.edu}
\affiliation{Department of Astronomy, University of Illinois at
  Urbana-Champaign, 1002 W.~Green Street, Urbana, IL 61801 USA}
\author{Benjamin D. Wandelt}
\email{bwandelt@uiuc.edu}
\affiliation{Institut d'Astrophysique de Paris,
Universite Pierre et Marie Curie (Paris 6),
98 bis, boulevard Arago,
75014 Paris,
France\\
Department of Astronomy, University of Illinois at
  Urbana-Champaign, 1002 W.~Green Street, Urbana, IL 61801 USA
}

\date{\today}
\maketitle 
Our calculations and results in \cite{kw07} are  correct. That we have to measure the
21-cm brightness temperature $T_b$ with cosmic microwave background (CMB) temperature $T_{CMB}$ as a
standard  is implicit in the definition of the 21-cm brightness
temperature. The 21-cm brightness temperature is defined as the difference of
the observed temperature (which is a function of frequency) and the CMB
temperature (which is independent of frequency). Thus we agree that the
quantity to be measured is $T_b/T_{CMB}$ but this has no effect on any of our
 calculations or conclusions in \cite{kw07}.

 Before
proceeding let us remark that in any  theory with dimensionful
quantities we have to fix a few conversion factors between different
dimensions, for example, the speed of light $c$, Boltzmann's constant $k_B$,
Planck's constant $h$ etc. We will fix these conversion factors and measure
all quantities in the same physical units.

Cosmological measurements are different from lab or quasar measurements
because in cosmology we can find standards which remain invariant  with respect
to the variation of microphysical parameters like the fine structure
constant on time scales of the order of the age of the Universe. In the
present case of 21-cm cosmology the relevant standard is provided by the
CMB temperature. The CMB temperature and the baryon to photon ratio
$\eta=n_B/n_{\gamma}$, where $n_B$ is the baryon number density and
$n_{\gamma}$ is the photon number density is fixed (apart from negligible
changes due to scattering  at later times) when the temperature of the
Universe is around $0.5\rm{MeV}$ and the electrons and positrons have
annihilated. In addition the relative number density of protons or Hydrogen
$n_H$ is fixed once the big bang nucleosynthesis is completed around
$T=0.01\rm{MeV}$. We reiterate that $\eta$, $T_{CMB}$ and $n_H$ are strict
constants after recombination, redshift $z\lesssim 1100$, in standard
cosmology apart from trivial evolution due to the expansion of the
Universe. In particular these are not affected by the variation in the fine
structure constant $\alpha$ or other microphysical constants. Although $n_H$ is
 changed by processing by stars at late times, this does not affect our
 calculations which depend only on  $n_H$ during the dark ages and which can be
 constrained using the 21-cm signal.

Any data analysis constraining $\alpha$ will have to marginalize over
the cosmological parameters including the primordial $n_H$ and $\eta$. These parameters
do not depend on alpha since we don't derive them from any fundamental
physics but use them as free parameters to be determined by a joint fit of cosmological
observatons including 21-cm observations of the dark ages, CMB and big bang
nucleosynthesis (BBN). Once
fixed during very early Universe these parameters are independent of
$\alpha$ although we do not know their exact values. The unique  frequency
dependence of the signal due to variation in $\alpha$ means that we should
not expect significant degeneracies with other cosmological parameters. We
agree (and mention in \cite{kw07}) that extending our work to variations of
other fundamental constants would be of interest.

Thus we can think of measuring the 21-cm transition frequency $\nu_{21}$
(and also the electron mass $m_e$) in
the units of CMB temperature which gives $\nu_{21}\propto \alpha^4$ used in
\cite{kw07}. This is in fact what physically also happens. The absorption
of the CMB photons by neutral hydrogen amounts to a measurement of
$\nu_{21}$ with respect to the CMB blackbody spectrum. Note that the same arguments cannot be used
for intensity of 21-cm radiation $I_{\nu_{21}}$ since the intensity of the
CMB radiation is also a function of frequency and not a strict constant
as the CMB temperature, i.e. the CMB intensity at  $\nu_{21}$ depends on $\nu_{21}$.

Similar arguments apply to the comments by Flambaum and Porsev \cite{fp} following
their Equation 2. In their dimensionless parameter $X_H$, $\eta$ is a fixed
constant after recombination. Although we do not know the origin or the
exact value of this constant, it is nevertheless indepedent of $\alpha$ or
any other microphysics at $z\lesssim 1100$. We can form another
dimensionless quantity $\nu_{21}/T_{CMB}$ and we have $X_H\propto
T_{CMB}/\nu_{21}\propto \alpha^{-4}$ which is consistent with the
calculation in \cite{kw07}.  We have already taken the effect of $\alpha$ on
recombination and hence on the ionization fraction  $x_e$ into account in
\cite{kw07} following previous CMB calculations. The change in $x_e$ of
course affects the spin temperature $T_S$ and hence the 21-cm signal. The
effect of recombination on the $\eta$ is negligible for our purpose as has
been shown in the recent precision calculations of recombination (with
changes in $T_{CMB}\sim nK$ compared to $\sim mK$ 21-cm signal). We remark again that all quantities in $X_H$
except $\nu_{21}$ are fixed by cosmology and independent of microphysics,
in particular $\alpha$, during and after recombination at $z\lesssim
1100$. The arguments of Flambaum and Porsev \cite{fp}  about
ignoring $X_H$ are thus incorrect.

\end{document}